\documentclass[onecolumn,preprint,showpacs,superscriptaddress]{revtex4-1}

\usepackage[utf8]{inputenc}
\usepackage{mathtools}
\usepackage{graphicx}
\usepackage{caption}
\usepackage{subcaption}
\usepackage{xcolor}
\usepackage{soul}
\usepackage{float}
\usepackage{amsmath}
\usepackage{amssymb}
\usepackage{amsfonts}
\usepackage{orcidlink}
\usepackage{appendix}

\begin{document}

\title{Kiselev black strings in $f(R,T)$ gravity}
%\title{Charged and Rotating Black String of Kiselev and Hawking Radiation from Tunneling of Scalar Particles}

\author{L. C. N. Santos \orcidlink{0000-0002-6129-1820}}
\email{luis.santos@ufsc.br}
\affiliation{Departamento de Física, CFM - Universidade Federal de \\ Santa Catarina; C.P. 476, CEP 88.040-900, Florianópolis, SC, Brazil}

\author{L. G. Barbosa  \orcidlink{0009-0007-3468-3718}}
\email{leonardo.barbosa@posgrad.ufsc.br}
\affiliation{Departamento de Física, CFM - Universidade Federal de \\ Santa Catarina; C.P. 476, CEP 88.040-900, Florianópolis, SC, Brazil}

\author{C. C. Barros Jr. \orcidlink{0000-0003-2662-1844}}
\email{barros.celso@ufsc.br }
\affiliation{Departamento de Física, CFM - Universidade Federal de \\ Santa Catarina; C.P. 476, CEP 88.040-900, Florianópolis, SC, Brazil}

\begin{abstract}
In this work, we investigate exact black string solutions in the context of $f(R,T)$ gravity. Adopting the specific form $f(R,T) = R + 2\chi T$, we consider an anisotropic Kiselev fluid as the matter content and obtain static cylindrical solutions, which are then extended to the rotating case through a suitable coordinate transformation. The influence of the quintessence state parameter $w_q$ and the matter--geometry coupling constant $\chi$ on the geometry is analyzed. We examine the weak, null, and strong energy conditions, identifying the regions in the parameter space where they are satisfied. Furthermore, we apply the Hamilton--Jacobi method to study the tunneling of scalar particles across the event horizon and derive the corresponding Hawking temperature. The thermodynamic stability of the solutions is investigated by computing the heat capacity, and the conditions for phase transitions are discussed. The results provide a characterization of black strings in $f(R,T)$ gravity surrounded by quintessence, highlighting the combined effects of anisotropic matter and modified gravity on their physical properties.

\end{abstract}

\maketitle

\section{Introduction}
 
Black strings, as proposed by Lemos~\cite{Lemos:1994xp}, are exact solutions of general relativity in the presence of a negative cosmological constant (anti-de Sitter) that describe cylindrical black holes. These objects possess intriguing geometric and thermodynamic properties. In contrast to usual black holes, whose horizons are compact, black strings can extend infinitely along a spatial dimension, leading to a topology of the form $\mathbb{S} \times \mathbb{R}$.

Although cylindrically symmetric solutions such as cosmic strings \cite{Vilenkin:1984ib} have been known for some time, solutions describing the cylindrical analogue of a black hole have been proposed only more recently \cite{Lemos:1994xp}. Such cylindrical objects can be obtained within the framework of general relativity and can be related to solutions in the context of three-dimensional dilaton--gravity theory \cite{Horowitz1993PhRvL}. In the context of black string solutions, the cosmological constant plays a fundamental role in shaping spacetime geometry and horizon structure. Motivated by the foundational work on such cylindrical objects, a wide range of solutions and studies has emerged, addressing both their geometric characteristics and thermodynamic behavior~\cite{Ahmed:2025sav,Shaikh:2025vmw,Barbosa:2025scy,Moughal:2025axt,Deglmann:2025mcl,Bakhtiarizadeh:2025uks,Pereira:2024cfa,Ahmed:2024gdj,Lessa:2024gbd,Darlla:2024bsv,Mirekhtiary:2024dmy}. The collapse of a cylindrically symmetric system has been suggested as a useful framework to represent a prolate collapse~\cite{Thorne:1972ji}, which may exhibit physical features distinct from those found in the usual spherical scenario. Such cylindrically symmetric models stand in contrast to the hoop conjecture \cite{Thorne:1972ji}, which asserts that a black hole forms from a massive star only when it is compressed into a region whose circumference is limited in every spatial direction. It was subsequently pointed out~\cite{Lemos:1995cm} that this conjecture was established under the assumption of a spacetime with no cosmological constant. Introducing this constant into the field equations leads to solutions with cylindrical symmetry that can correspond to either black holes or black strings.

Recently, modifications to the field equations derived from general relativity have been proposed, motivated by challenges as quantization of the gravitational field as well as by the difficulties in explaining the nature of dark matter and dark energy in the Universe. In this context, a direct modification to the gravitational action \cite{fdr2} consists in replacing the Ricci scalar curvature $R$ in the geometric sector by a general function $f(R, T)$, where $T$ denotes the trace of the energy--momentum tensor $T_{\mu\nu}$. Within the framework of this gravitational theory, the existence of spherically symmetric solutions describing black holes surrounded by a Kiselev fluid has been investigated more recently \cite{santos2023kiselev}. This type of anisotropic fluid has also been the subject of increasing attention in the recent literature considering  shadow of black holes~\cite{shadow1,shadow2,shadow3,shadow4,shadow5}, quasinormal modes~\cite{quasi1,quasi2,quasi3,quasi4,quasi5,quasi6,quasi7,quasi8,quasi9,quasi10,quasi11} and  thermodynamics of black holes~\cite{termo1,termo2,termo3,termo4,termo5,termo6,termo7,termo8,termo9,termo10,termo11,termo12,termo13,termo14,termo15}.

In this work, we consider the field equations in $f(R, T)$ gravity and an anisotropic Kiselev fluid \cite{Kiselev:2002dx} and derive the corresponding black string solution, including its rotating version. We analyze the energy conditions for the obtained spacetime and identify the parameter ranges where they are satisfied. We proceed to study scalar particle tunneling from the rotating black string, deriving the Hawking temperature and the associated tunneling rate. We also investigate the heat capacity of the solution.

In Section \ref{sec2}, we review the field equations in $f(R,T)$ gravity focusing on the case $f(R,T) = R + 2\chi T$.  
In Section \ref{Solution_of_the_field_equation}, we obtain black string solutions immersed in a Kiselev fluid and extend them to the rotating case.  
In Section \ref{energycond}, we analyze the corresponding energy conditions and determine the regions where they are satisfied.  
In Section \ref{tunneling}, we study scalar particle tunneling and derive the Hawking temperature associated with the rotating black string.  
In Section \ref{heat}, we compute the heat capacity and discuss the thermodynamic stability of the solutions.  
In Section \ref{conclusion}, we present the main conclusions and summarize the results.

\section{Field equations in $R + 2\chi T$ gravity} \label{sec2}

In this section, we provide a concise overview of the field equations within the framework of $f(R,T)$ gravity \cite{fdr2}. In this theory, the action is expressed as
\begin{equation}
S=\frac{1}{16\pi}\int f(R,T)\sqrt{-g} \, d^4x + \int L_{m}\sqrt{-g} \, d^4x,
    \label{e1}
\end{equation}
where $f(R,T)$ denotes a general function depending on the Ricci scalar $R$ and the trace $T$ of the matter energy-momentum tensor. Here, $L_m$ in Eq.~(\ref{e1}) represents the matter Lagrangian density, which determines a specific form for the energy-momentum tensor. By performing a variation of the action $S$ with respect to the metric tensor, we obtain the following expression:

\begin{align}
\delta S=& \frac{1}{16\pi}\int\left[f_{R}(R,T)R_{\mu\nu}\delta g^{\mu\nu} + f_{R}(R,T)g_{\mu\nu} \Box \delta  g^{\mu\nu}  \right.  
\nonumber\\
& -f_{R}(R,T)\nabla_{\mu}\nabla_{\nu}\delta g^{\mu\nu} + 
 f_{T}(R,T)\frac{\delta (g^{\eta\xi}T_{\eta\xi})}{\delta g^{\mu\nu}}\delta g^{\mu\nu}
\nonumber \\
 & \left. -\frac{1}{2}g_{\mu\nu}f(R,T)\delta g^{\mu\nu}+\frac{16\pi}{\sqrt{-g}}\frac{\delta (\sqrt{-g}L_{m})}{\delta g^{\mu\nu}}\right]\sqrt{-g} \, d^4x,  
    \label{e2}
\end{align}
where we have introduced $f_R(R,T)=\partial f(R,T)/\partial R$ and $f_T(R,T)=\partial f(R,T)/\partial T$ to denote the partial derivatives of $f(R,T)$ with respect to $R$ and $T$, respectively. After integrating the second and third terms and incorporating the variation of $T$, we obtain
\begin{equation}
\frac{\delta (g^{\eta\xi}T_{\eta\xi})}{\delta g^{\mu\nu}}=T_{\mu\nu} + \Theta_{\mu\nu}
    \label{e3},
\end{equation}
with
\begin{equation}
\Theta_{\mu\nu} \equiv g^{\eta\xi}\frac{\delta T_{\eta\xi}}{\delta g^{\mu\nu}}=-2T_{\mu\nu}+g_{\mu\nu}L_{m}-2g^{\eta\xi}\frac{\partial^2L_{m}}{\partial g^{\mu\nu}\partial g^{\eta\xi}},
    \label{e4}
\end{equation}
resulting from the definition of the variation of the matter Lagrangian density. As a consequence, the field equations take the form
\begin{align}
 f_{R}&(R,T)R_{\mu\nu}-\frac{g_{\mu\nu}}{2}f(R,T)+
(g_{\mu\nu}\Box - \nabla_{\mu}\nabla_{\nu})f_{R}(R,T) \nonumber\\
&= 8\pi T_{\mu\nu} - f_{T}(R,T)T_{\mu\nu}- f_{T}(R,T)\Theta_{\mu\nu}\, ,
   \label{e5}
\end{align}
which represents the version of the field equation in $f(R,T)$ gravity that will be employed in this work. A commonly explored form of the function $f(R,T)$ is given in Ref.~\cite{fdr2}:
\begin{equation}
    f(R,T)=R + 2f(T),
    \label{e6}
\end{equation}
where $f(T)$ stands for an arbitrary function depending on the trace of the energy-momentum tensor. Substituting Eq.~(\ref{e6}) into Eq.~(\ref{e5}), the field equations are rewritten as
\begin{align}
 R_{\mu\nu}-\frac{g_{\mu\nu}}{2}R=& 8\pi T_{\mu\nu}- 2f'(T)T_{\mu\nu}
 \nonumber\\
 &- 2f'(T)\Theta_{\mu\nu} + f(T)g_{\mu\nu},
    \label{e7}
\end{align}
where $f'(T)=df(T)/dT$ denotes the derivative of $f$ with respect to $T$. In the presence of the cosmological constant 
$\Lambda=-\frac{3}{\ell^{2}}$, where $\ell$ is the AdS radius, we can write
\begin{align}
 R_{\mu\nu}-\frac{g_{\mu\nu}}{2}R-\Lambda g_{\mu\nu}=& 8\pi T_{\mu\nu}- 2f'(T)T_{\mu\nu}
 \nonumber\\
 &- 2f'(T)\Theta_{\mu\nu} + f(T)g_{\mu\nu}.
    \label{e7}
\end{align}
Thus, this field equation can be explicitly written with a particular choice of $f(T)$. For an anisotropic fluid, we may choose the Lagrangian density function as $L_m = (p_r + 2p_t)/3$ \cite{Deb:2018sgt}, where $p_r$ and $p_t$ are the radial and tangential pressures, respectively. In this work, we consider the function $f(T)$ to by gives as $f(T)=\chi T$. In the next section, we will consider a particular equation of state for $T_{\mu}^{\nu}$ and explicitly derive the associated differential equations.

\section{Solution of the field equation}\label{Solution_of_the_field_equation}

%%%%%%%%%%%%%%%%%%%%%%%%%%%%%%%%%%%%%%%%%%%%%%
\begin{figure}[!ht]
\centering
\includegraphics[width=\textwidth,height=\textheight,keepaspectratio]{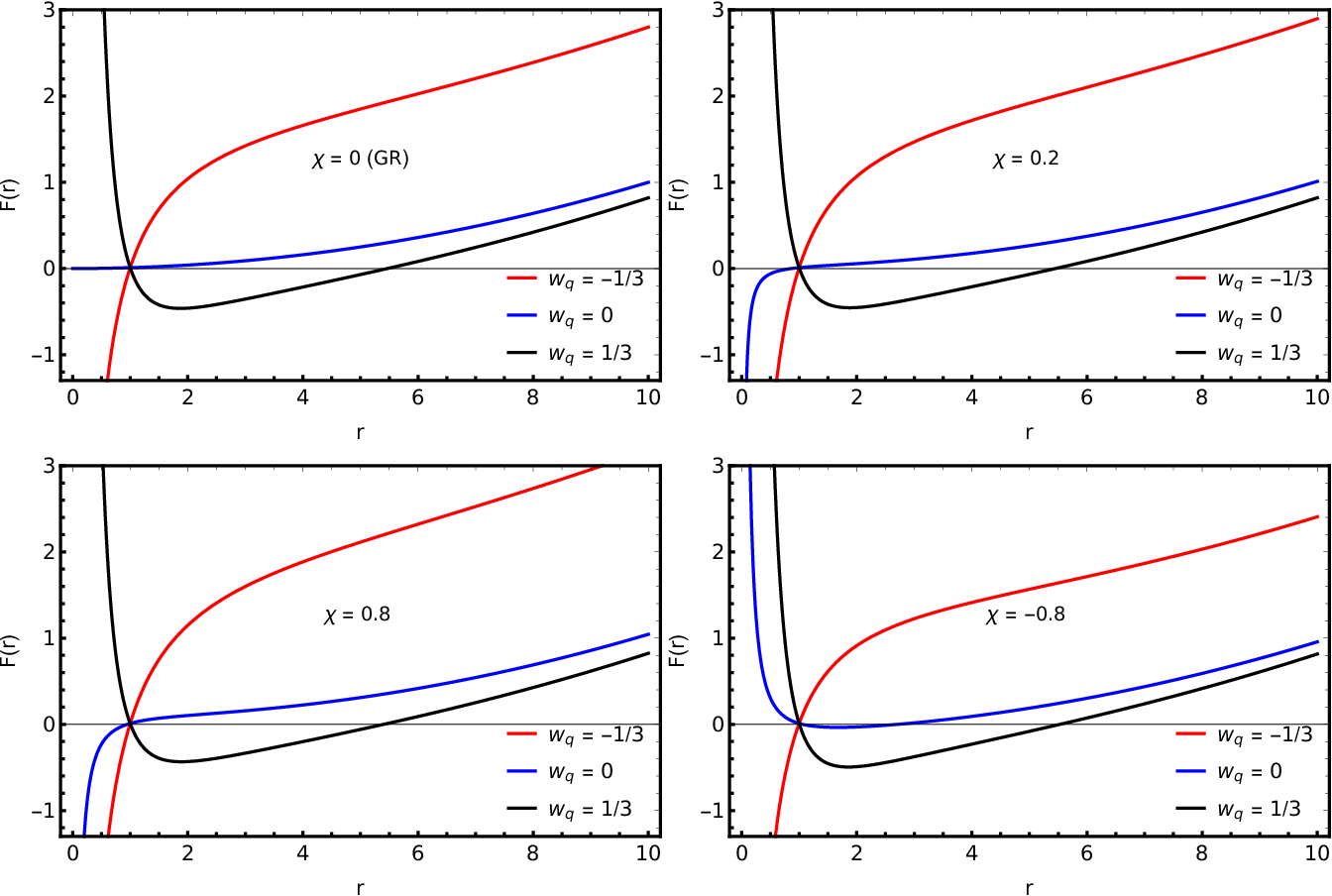} 
\caption{Behavior of the metric function $F(r)$ as a function of the radial coordinate $r$. The plots are shown for fixed parameters $K = 2$, $\ell = 10$, and $M = 1$, and different values of the quintessence state parameter $w_q$ and the coupling parameter $\chi$. Each panel corresponds to a different value of $\chi$, namely $\chi = 0$, $0.2$, $0.8$, and $-0.8$. The curves represent $w_q = -1/3$ (red), $w_q = 0$ (blue), and $w_q = 1/3$ (black).}
\label{fig1}
\end{figure}
%%%%%%%%%%%%%%%%%%%%%%%%%%%%%%%%%%%%%%%%%%
For the version of $f(R,T)$ gravity we explore in this work, it is important to consider the presence of matter in order to obtain a nontrivial structure on the right-hand side of the field equations. To this end, we study an effectively anisotropic fluid. Considering the Kiselev fluid \cite{Kiselev:2002dx}, which is characterized by the following components of the energy–momentum tensor
\begin{equation}
    T_{t}^{t}=T_{r}^{r}=\rho,\quad T_{\phi}^{\phi}=T_{z}^{z}=-\frac{1}{2}\rho\left(3w_q+1\right).
    \label{Tkiselev}
\end{equation}
Starting from the line element for a general static black string in cylindrical coordinates \cite{Lemos:1994xp}:
\begin{equation}
   ds^{2}=F\left(r\right)dt^{2}-\frac{dr^{2}}{F\left(r\right)}-r^{2}d\phi^{2}-\frac{r^{2}}{\ell^{2}}dz^{2}, \label{bsStatic}
\end{equation}
where $-\infty<t<\infty$, $0\leq r<\infty$, $0\leq \phi<2\pi$ and $-\infty<z<\infty$. Calculating the $f(R,T)$ field equations, Eq.~(\ref{e7}), using the line element for a static black string, Eq.~(\ref{bsStatic}), and the energy-momentum tensor for a Kiselev fluid, Eq.~(\ref{Tkiselev}), we obtain the expressions:
\begin{equation}
   \frac{1}{r}F^{'}\left(r\right)+\frac{1}{r^{2}}F\left(r\right)-\frac{3}{\ell^{2}}=-\alpha\rho\left(r\right), \label{field1}
\end{equation}
and
\begin{equation}
  \frac{1}{2}F^{''}\left(r\right)+\frac{1}{r}F^{'}\left(r\right)-\frac{3}{\ell^{2}}=\beta\rho\left(r\right), \label{field2}
\end{equation}
where the prime denotes differentiation with respect to the radial coordinate. 
We also define 
$\alpha = 8\pi + \left(3-w_{q}\right)\chi$  and  $\beta = 4\pi\left(3w_{q}+1\right) + 4\chi w_{q}$.  Combining Eqs. (\ref{field1}) and (\ref{field2}), we get the following differential equation
\begin{equation}
    \frac{\alpha}{2}F^{''}\left(r\right)+\frac{\left(\alpha+\beta\right)}{r}F^{'}\left(r\right)+\frac{\beta}{r^{2}}F\left(r\right)-\left(\alpha+\beta\right)\frac{3}{\ell^{2}}=0,
\end{equation}
which admits an exact solution given by
\begin{equation}
 F\left(r\right)=\frac{r^{2}}{\ell^{2}}-\frac{2M}{r}+\frac{K}{r^{n_{w_{q},\chi}}},
    \label{eq14}
\end{equation}
where $M$ and $K$ are constants of integration and, in which we define $n_{w_{q},\chi}=\frac{2\beta}{\alpha}$. So that the line element assumes
\begin{equation}
    ds^{2}=\left(\frac{r^{2}}{\ell^{2}}-\frac{2M}{r}+\frac{K}{r^{n_{w_{q},\chi}}}\right)dt^{2}-\frac{dr^{2}}{\left(\frac{r^{2}}{\ell^{2}}-\frac{2M}{r}+\frac{K}{r^{n_{w_{q},\chi}}}\right)}-r^{2}d\phi^{2}-\frac{r^{2}}{\ell^{2}}dz^{2}, \label{dsStatic}
\end{equation}
which characterizes the solution of a static black string immersed in the Kiselev fluid within 
$f(T,R)=R+2\chi T$ gravity. The equation governing the structure of the event horizon is given by  
\begin{equation}
    \ell^{2} r^{\,n_{w_{q},\chi}+3} - 2M \ell^{2} r^{\,n_{w_{q},\chi}} + \ell^{2} K r = 0 \, .
\end{equation}
The existence of solutions to this equation can be investigated by analyzing the behavior of $F(r)$ graphically. The roots of this function, denoted by $r_{+}$, correspond to the event horizons of the spacetime. 

In Fig. \ref{fig1}, we show the behavior of the metric function $F(r)$ given by Eq.~(\ref{eq14}) as a function of the radial coordinate $r$ for different values of the quintessence and coupling parameters $w_{q}$ and $\chi$, respectively. In the top left panel, we show $F(r)$ for the GR case, we can observe that, for the values of $K$, $l$ and $M$ considered in this plot, each value of the quintessence parameter leads to a distinct behavior. The curve for $w_{q}=-1/3$ presents one horizon, it takes negative values for $r$ close to the origin, crosses the axis once, and then increases rapidly for larger $r$. In contrast, for $w_{q}=0$, $F(r)$ approaches zero from positive values but never crosses the axis, indicating that this solution do not present a horizon. As for the case $w_{q}=1/3$, the metric function crosses the axis two times, which can be associated with the presence to two horizons. In the top right panel, we present the same three quintessence cases, now in the context of $f(R,T)$ gravity with $\chi=0.2$. Analyzing this figure, we can conclude that, for this value of $\chi$, the modifications in the behavior of the metric function are minimal. However, a noticeable difference can be observed for the solution with $w_q=0$: now the curve crosses the axis to negative values, indicating the presence of a horizon in $f(R,T)$ gravity, whereas no horizon was present in the corresponding GR case. In the bottom left panel, we increase the value of the coupling of the modified gravity $(\chi=0.8)$, and it is possible to notice that all curves grow to positive values more rapidly. On the other hand, when we decrease the value of $\chi$ to $-0.8$, in the bottom right panel, the opposite behavior can be observed. That is, in this case all curves increase more slowly to positive values. At the same time, for $\chi=-0.8$, the behavior of the quintessence curve with $w_{q}=0$ is modified once again, this time it crosses the axis two times, indicating the presence of two horizons.

We can obtain a solution for a rotating black string following Lemos' work
\begin{equation}
    t\rightarrow\lambda t-a\phi,\quad\phi\rightarrow\lambda\phi-\frac{a}{\ell^{2}}t
\end{equation}
where $ \lambda=\sqrt{1+\frac{a^{2}}{l^{2}}}$, which results in the following line element
\begin{equation}\label{eq11}
    ds^{2}=\left(\frac{r^{2}}{\ell^{2}}-\frac{2M}{r}+\frac{K}{r^{n_{w_{q},\chi}}}\right)\left(\lambda dt-ad\phi\right)^{2}-\frac{dr^{2}}{\left(\frac{r^{2}}{\ell^{2}}-\frac{2M}{r}+\frac{K}{r^{n_{w_{q},\chi}}}\right)}-r^{2}\left(\lambda d\phi-\frac{a}{\ell^{2}}dt\right)^{2}-\frac{r^{2}}{\ell^{2}}dz^{2},
\end{equation}
Which represents a rotational solution of its previously obtained static counterpart, Eq.~(\ref{dsStatic}), so that when $a=0$ we recover the static solution.

\section{Energy conditions} \label{energycond}

%%%%%%%%%%%%%%%%%%%%%%%%%%%%%%%%%%%%%%%%%%%%%%
\begin{figure}[!ht]
\centering
\includegraphics[scale=0.7]{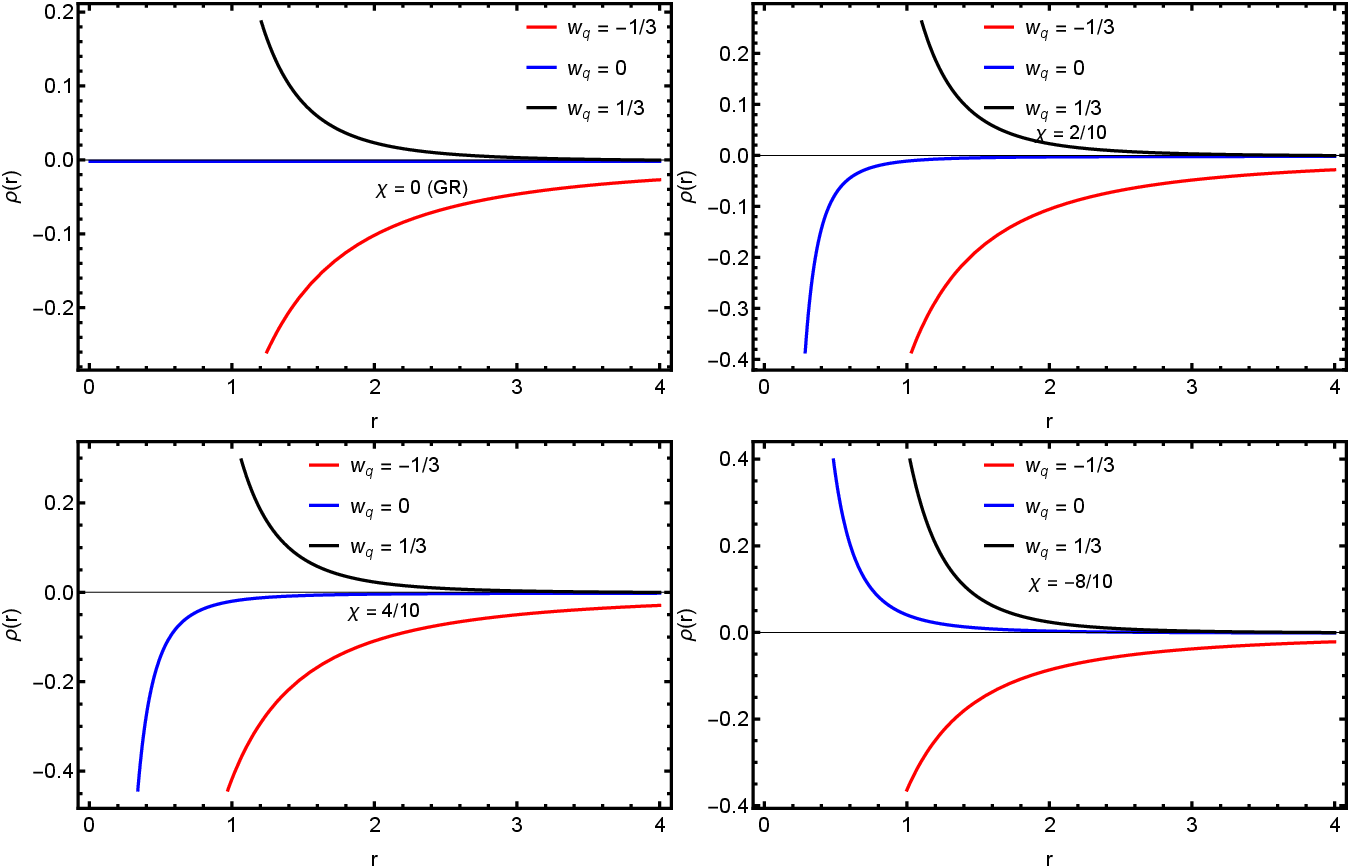} 
\caption{Energy density $\rho(r)$ as a function of the radial coordinate $r$ for a black string configuration. The plots are shown for fixed parameters $K = 0.4$, $\ell = 40$, and $M = 1$, and different values of the quintessence state parameter $w_q$ and the coupling parameter $\chi$. Each panel corresponds to a different value of $\chi$, namely $\chi = 0$, $0.2$, $0.4$, and $-0.8$. The curves represent $w_q = -1/3$ (red), $w_q = 0$ (blue), and $w_q = 1/3$ (black).
}
\label{fig2}
\end{figure}
%%%%%%%%%%%%%%%%%%%%%%%%%%%%%%%%%%%%%%%%%%

In this section, we consider the energy conditions associated with a spacetime represented by the line element from Eq. (\ref{eq11}). We now proceed to define a tetrad basis such that the energy–momentum tensor is diagonal:
\begin{equation}
    e^{(a)}_{\:\:\:\:\mu}=\left(
\begin{array}{cccc}
\lambda \sqrt{F(r)} & 0 & -a \sqrt{F(r)} & 0 \\
0 & \dfrac{1}{\sqrt{F(r)}} & 0 & 0 \\
\dfrac{a r}{\ell^2} & 0 & -r \lambda & 0 \\
0 & 0 & 0 & -\dfrac{r}{\ell}
\end{array}
\right),
\label{eq19}
\end{equation}
where $ e^{(a)}_{\:\:\:\:\mu}$ satisfies the relation $g_{\mu\nu}=  e^{(a)}_{\:\:\:\:\mu} e^{(b)}_{\:\:\:\:\nu}\eta_{(a)(b)}$ with $\eta_{(a)(b)}$ being the usual Minkowski metric. By using Eq. (\ref{eq19}), the energy--momentum tensor is given by the expression 
\begin{equation}
    T^{(a)(b)} = \frac{1}{8\pi}e^{(a)}_{\:\:\:\:\mu} e^{(b)}_{\:\:\:\:\nu}G^{\mu\nu}.
\end{equation}
Based on the chosen tetrad basis and using the definition of the Einstein tensor, the components of effective energy density and pressures are given in the form
\begin{align}
\rho_{eff}&=T^{(t)}_{\:\:\:\:(t)}=-\frac{1}{8\pi}\left(\frac{1}{r} \frac{d F(r)}{dr} + \frac{F(r)}{r^2}
\right),\\
p_r &= -T^{(r)}_{\:\:\:\:(r)}=\frac{1}{8\pi}\left(\frac{1}{r} \frac{d F(r)}{dr} + \frac{F(r)}{r^2}
\right),\\
p_t &=-T^{(\phi)}_{\:\:\:\:(\phi)}=-T^{(z)}_{\:\:\:\:(z)}=\frac{1}{8\pi}\left(\frac{1}{2} \frac{d^2 F(r)}{dr^2} + \frac{1}{r} \frac{d F(r)}{dr}\right).
\end{align}

We can see that the energy density, as well as the radial and tangential pressures, depend on the metric function obtained in Eq. (\ref{eq14}). It is straightforward to verify that the sum of the energy density and the radial pressure vanishes. Consequently, any energy conditions involving the inequality $\rho_{eff}+p_r \geq0$ are trivially satisfied.    

%%%%%%%%%%%%%%%%%%%%%%%%%%%%%%%%%%%%%%%%%%%%%%
\begin{figure}[!ht]
\centering
\includegraphics[scale=1]{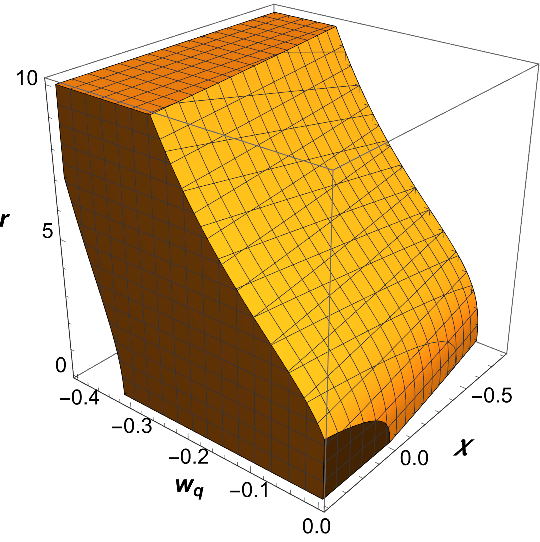} 
\caption{This is the plot of the region where all the energy conditions considered in this paper are satisfied, for $M=1$, $l=40$, and $K=-0.4$.
}
\label{fig3}
\end{figure}
%%%%%%%%%%%%%%%%%%%%%%%%%%%%%%%%%%%%%%%%%%

We shall now apply these results to examine the following energy conditions:

\begin{align}
\text{Weak Energy Condition (WEC): }& ~\rho_{eff} \geqslant 0,~\rho_{eff} + p_i \geqslant 0
\label{eq12}\\
\text{Null Energy Condition (NEC): }& ~\rho_{eff} + p_i \geqslant 0  \label{eq13}\\
\text{Strong Energy Condition (SEC): }& \rho_{eff} + p_i \geqslant 0, ~ \rho_{eff} +\sum p_i\geqslant 0
\label{eq26}
\end{align}

In Fig. \ref{fig2}, it is possible that the case $w_q = 0$  is more strongly affected by $\chi$ and consequently the energy density, as a function of the radial coordinate, exhibits a distinct behavior in each of the plots shown in the figure. In Fig.~\ref{fig3}, we illustrate the region where all the energy conditions mentioned above are simultaneously satisfied for fixed values of the parameters $M$, $l$ and $K$. We can observe that, for the set of parameter values analyzed in Fig.~\ref{fig3}, only negative values of Kiselev parameter $w_q$ satisfy all energy conditions, while there exists a region involving both positive and negative values of $\chi$ that satisfies the conditions.

%\section{Particular Cases}
%\subsection{Charged black string}
%\subsection{Pure Lemos black string in $f(R,T)$}
%\subsection{Regular solution?}

\section{Scalar particle tunneling} \label{tunneling}

To study the contribution of scalar particles to the Hawking radiation of rotating black strings immersed in an anisotropic Kiselev fluid in gravity $f(R,T)$, we consider the Klein-Gordon equation written in the metric derived in section (\ref{Solution_of_the_field_equation}),
\begin{equation}
    \frac{1}{\sqrt{-g}}\partial_{\alpha}\left(\sqrt{-g}g^{\alpha\beta}\partial_{\beta}\Psi\right)-\frac{m^{2}}{\hslash^{2}}\Psi=0
\end{equation}
where $m$ is the mass of the scalar particles. We apply the WKB approximation and assume an analysis of the form
\begin{equation}
    \Psi\left(t,r,\phi,z\right)=\exp\left[\frac{i}{\hslash}I\left(t,r,\chi,z\right)\right],
\end{equation}
to leading order in $\hbar$, this yields the relativistic Hamilton--Jacobi equation.
\begin{equation}
    g^{\alpha\beta}\left(\partial_{\alpha}I\partial_{\beta}I\right)+m^{2}=0
\end{equation}

Using the coefficients from Eq.~(\ref{DiagMetric}) and simplifying, we obtain
\begin{equation}
g^{tt}\left(\partial_{t}I\right)^{2}+g^{rr}\left(\partial_{r}I\right)^{2}+g^{\chi\chi}\left(\partial_{\chi}I\right)^{2}+g^{zz}\left(\partial_{z}I\right)^{2}+m^{2}=0
\end{equation}

Taking into account the spacetime symmetries of the background metric, we assume the solution to the above equation can be written as
\begin{equation}
I = -\bigl(E - J_1 \Omega_+\bigr) t + W(r) + J_1 \chi + J_2 z,
\end{equation}
where $E$, $J_1$, and $J_2$ denote the quantum numbers corresponding to the energy, angular momentum, and momentum along the $z$-direction, respectively. Here, $\Omega_+$ is the angular velocity evaluated at the event horizon. Substituting the quantities appropriately, we can write
\begin{equation}
    W_{\pm}\left(r\right)=\pm\int dr\frac{\sqrt{\left[\lambda^{2}-\frac{a^{2}}{r^{2}}F\left(r\right)\right]\left[\left(E-J_{1}\Omega\right)^{2}-\frac{A\left(r\right)}{C\left(r\right)}J_{1}^{2}+A\left(r\right)\left(g^{zz}J_{2}^{2}+m^{2}\right)\right]}}{F\left(r\right)}. 
\end{equation}

%%%%%%%%%%%%%%%%%%%%%%

By applying the residue theorem to evaluate the integral in the action, we obtain the imaginary part of the principal function for the outgoing (\(+\)) and ingoing (\(-\)) particles at the horizon:
\begin{equation}
W_{\pm}(r) = \pm \frac{i\pi\lambda (E - J_1 \Omega)}{F'(r_+)}.
\label{eq:Wpm}
\end{equation}
The tunneling probability for the particle to cross the horizon is given by $\Gamma \propto \exp[-2\,\text{Im}\,W]$. Using the result for the outgoing particle ($W_+$), the ratio of the emission-to-absorption probabilities is therefore
\begin{equation}
\Gamma \equiv \frac{P_{\text{emission}}}{P_{\text{absorption}}} = \exp\left[-4\,\text{Im}\,W_+\right] = \exp\left[-\frac{4\pi \lambda (E - J_1 \Omega)}{F'(r_+)}\right].
\label{eq:GammaGeneral}
\end{equation}

This takes the form of a Boltzmann factor $\Gamma = \exp[-(E - J_1\Omega)/T_H]$, from which we can immediately identify the Hawking temperature of the black hole as
\begin{equation}
T_{\text{H}} = \frac{F'(r_+)}{4\pi \lambda}.
\label{eq:TempGeneral}
\end{equation}

For the specific case under consideration, the function $F(r)$ is defined such that
\begin{equation}
F'(r_+) = \frac{2r_+}{\ell^2} + \frac{2M}{r_+^2} - \frac{K\,n_{w_q,\chi}}{r_+^{n_{w_q,\chi}+1}}.
\end{equation}
Substituting this expression into Eqs.~(\ref{eq:GammaGeneral}) and (\ref{eq:TempGeneral}) yields the explicit results for the tunneling rate and Hawking temperature:
\begin{equation}\label{eq:GammaSpecific}
   \Gamma=\exp\left[-4\pi\lambda\left(E-J_{1}\Omega\right)\left(\frac{2r_{+}}{\ell^{2}}+\frac{2M}{r_{+}^{2}}-\frac{Kn_{w_{q},\chi}}{r_{+}^{n_{w_{q},\chi}+1}}\right)^{-1}\right],
\end{equation}
\begin{equation}\label{TempSpecific}
T_{\text{H}}=\frac{1}{4\pi\lambda}\left(\frac{2r_{+}}{\ell^{2}}+\frac{2M}{r_{+}^{2}}-\frac{Kn_{w_{q},\chi}}{r_{+}^{n_{w_{q},\chi}+1}}\right).
\end{equation}

We can note that Eq.~(\ref{TempSpecific}) depends explicitly on the solution parameters, such as the event horizon radius $r_+$, the AdS radius $\ell$, the black string mass $M$, the anisotropic fluid parameters $K$ and $w_q$, and the modified gravity parameters $\chi$.

\section{Heat capacity} \label{heat}

We are interested in the heat capacity associated with the black-string solution. Recalling that \(F\left(r_{+}\right)=0\), we can isolate the mass of the black string
\begin{equation}
    M=\left(\frac{r_{+}^{2}}{\ell^{2}}+\frac{K}{r_{+}^{n_{w_{q},\chi}}}\right)\frac{r_{+}}{2}
\end{equation}
Substituting this expression into the temperature (Eq.~(\ref{TempSpecific})), we can rewrite the Hawking temperature as
\begin{equation}
    T_{\text{H}}=\frac{3}{4\pi\lambda r_{+}}\left(\frac{r_{+}^{2}}{\ell^{2}}+\frac{K\left(1-n_{w_{q},\chi}\right)}{3r_{+}^{n_{w_{q},\chi}}}\right)
\end{equation}
The thermodynamic stability of the solution is determined by the sign of the heat capacity: the black string is locally stable when the heat capacity is positive and unstable otherwise. The heat capacity at the horizon is defined by
\begin{equation}
    C_{+}=\left(\frac{dM}{dT}\right)_{r=r_{+}}
\end{equation}
Using the expressions for \(M\) and \(T_{\text{H}}\), we obtain
\begin{equation}
    C_{+}=\frac{2\pi\lambda r_{+}^{2}\left(\frac{r_{+}^{2}}{\ell^{2}}-\frac{\left(n_{w_{q},\chi}-1\right)K}{3r_{+}^{n_{w_{q},\chi}}}\right)}{\left(\frac{r_{+}^{2}}{\ell^{2}}+\frac{\left(1-n_{w_{q},\chi}\right)^{2}K}{3r_{+}^{n_{w_{q},\chi}-1}}\right)}
\end{equation}
For a given set of parameters there exists a critical radius \(r_{+}^{\text{c}}\) at which the denominator vanishes and \(C_{+}\rightarrow+\infty\). The equation that determines this critical radius is
\begin{equation}
    r_{\text{c}}^{n_{w_{q},\chi}+1}+\left(1-n_{w_{q},\chi}\right)^{2}\frac{\ell^{2}}{3}K=0
\end{equation}
whose solution is given by
\begin{equation}
    r_{+}^{\text{c}}=\left[-\left(1-n_{w_{q},\chi}\right)^{2}\frac{\ell^{2}}{3}K\right]^{\frac{1}{n_{w_{q},\chi}+1}}
\end{equation}
Note that a real, positive \(r_{+}^{\text{c}}\) exists only if the quantity inside the brackets is positive (for real \(n_{w_{q},\chi}\) this typically requires \(K<0\)).

\section{Conclusions} \label{conclusion}

In this work, we have obtained exact solutions describing static and rotating black strings in the framework of $f(R,T)$ gravity, with the matter content modeled as an anisotropic Kiselev fluid. By adopting the functional form $f(R,T) = R + 2f(T)$, we derived the metric functions analytically for arbitrary values of the quintessence state parameter $w_q$ and the matter--geometry coupling constant $\chi$. The resulting solutions generalize the Lemos black string to a modified gravity scenario with anisotropic matter fields. We have shown that the parameters $w_q$ and $\chi$ significantly affect the horizon structure, the behavior of the metric function, and the global properties of the spacetime.

The analysis of the energy conditions demonstrated that there exist well-defined regions in the spacetime where the weak, null, and strong energy conditions are simultaneously satisfied, including cases with negative values of $w_q$ and $\chi$. 

Using the Hamilton--Jacobi method, we studied the tunneling process of scalar particles across the event horizon and obtained an explicit expression for the Hawking temperature of the rotating black string. This analysis revealed the dependence of the temperature on both the quintessence parameter and the matter--geometry coupling, offering a thermodynamic characterization of the solutions. Furthermore, we computed the heat capacity at the horizon and identified critical points corresponding to divergences in $C_+$, which mark possible phase transitions between thermodynamically stable and unstable regimes. The stability analysis showed that the sign of the heat capacity is sensitive to the values of $w_q$ and $\chi$, thus linking the thermodynamic behavior directly to the matter and gravitational sector parameters.

Overall, our results extend the study of black strings to the context of $f(R,T)$ gravity, incorporating anisotropic fluids with quintessence-like behavior. The obtained solutions bridge modified gravity effects with cylindrical horizon topologies and provide a framework for exploring more general configurations. Future work may include the addition of electric charge, the study of quasinormal modes, and the investigation of observational signatures such as gravitational lensing and black string shadows in this extended gravitational setting.

\section{Acknowledgements}
LCNS would like to thank Conselho Nacional de Desenvolvimento Científico e Tecnológico - Brazil (CNPq) for financial support under Research Project No. 443769/2024-9 and Research Fellowship No. 314815/2025-2. LGB acknowledge the financial support from CAPES (process numbers 88887.968290/2024-00). LCNS is grateful to Franciele M. da Silva for insightful discussions and technical assistance.

\bibliographystyle{ieeetr}
\bibliography{sample}

\appendix
\section{Metric Transformation}\label{Metric_Transformation}
In this appendix, we present the metric transformation under the coordinate shift 
$\phi = \chi + \Omega t$. Starting from the general stationary line element
\begin{equation}
    ds^{2} = g_{tt}\,dt^{2} + g_{rr}\,dr^{2} + 2 g_{t\phi}\,dt\,d\phi 
    + g_{\phi\phi}\,d\phi^{2} + g_{zz}\,dz^{2},
\end{equation}
the change of variables leads to
\begin{equation}
    ds^{2} = (g_{tt} + 2\Omega g_{t\phi} + \Omega^{2} g_{\phi\phi})\,dt^{2} 
    + g_{rr}\,dr^{2} + 2 (g_{t\phi} + \Omega g_{\phi\phi})\,dt\,d\chi 
    + g_{\phi\phi}\,d\chi^{2} + g_{zz}\,dz^{2},
\end{equation}
imposing that the mixed term $dt\,d\chi$ is zero allows us to characterize $\Omega$ as follows
\begin{equation}
    \Omega = -\frac{g_{t\phi}}{g_{\phi\phi}}.
\end{equation}
Substituting this condition, the metric diagonalizes to
\begin{equation}
    ds^{2} = \left(g_{tt} - \frac{g_{t\phi}^{2}}{g_{\phi\phi}}\right)dt^{2} 
    + g_{rr}\,dr^{2} + g_{\phi\phi}\,d\chi^{2} + g_{zz}\,dz^{2}.
\end{equation}

For the black string solution immersed in an anisotropic quintessence fluid in $f(R,T)$, as obtained in section (\ref{Solution_of_the_field_equation})

\begin{equation}
\begin{split}
ds^{2} = & \left(F(r)\lambda^{2} - r^{2}\frac{a^{2}}{\ell^{4}}\right) dt^{2} 
- \frac{dr^{2}}{F(r)} 
- \left(F(r)\lambda a - r^{2}\lambda\frac{a}{\ell^{2}}\right) 2\,dt\,d\phi \\
& - \left(r^{2}\lambda^{2} - F(r)a^{2}\right) d\phi^{2} 
- \frac{r^{2}}{\ell^{2}}dz^{2} \, 
\end{split}
\end{equation}
where
\begin{equation}
   F\left(r\right)=\frac{r^{2}}{\ell^{2}}-\frac{2M}{r}+\frac{K}{r^{n_{w_{q},\chi}}},
\end{equation}
We identify the metric components as
\begin{equation}
\begin{aligned}
g_{tt} &= \left(F(r)\lambda^{2} - r^{2}\frac{a^{2}}{\ell^{4}}\right), \\
g_{rr} &= -\frac{1}{F(r)}, \\
g_{t\phi} &= g_{\phi t} = -\left(F(r)\lambda a - r^{2}\lambda\frac{a}{\ell^{2}}\right), \\
g_{\phi\phi} &= -\left(r^{2}\lambda^{2} - F(r)a^{2}\right), \\
g_{zz} &= -\frac{r^{2}}{\ell^{2}}.
\end{aligned}
\end{equation}

After the coordinate transformation, the metric adopts a diagonal form convenient for physical analysis:
\begin{equation}\label{DiagMetric}
    ds^{2} = A(r) \, dt^{2} - \frac{dr^{2}}{B(r)} - C(r) \, d\chi^{2} - D(r) \, dz^{2},
\end{equation}
where the metric functions are given by
\begin{equation}
\begin{aligned}
    A(r) &= \frac{F(r) \, r^{2}}{r^{2}\lambda^{2} - F(r)a^{2}}, \\
    B(r) &= F(r), \\
    C(r) &= r^{2}\lambda^{2} - F(r)a^{2}, \\
    D(r) &= \frac{r^{2}}{\ell^{2}}.
\end{aligned}
\end{equation}
The angular velocity \(\Omega\) associated with the spacetime rotation follows from the off-diagonal components of the original metric, yielding
\begin{equation}
\begin{aligned}
    \Omega(r) &= -\frac{F(r)\lambda a - r^{2}\lambda\, a/\ell^{2}}{r^{2}\lambda^{2} - F(r)a^{2}}, \\
    \Omega_{+} &= \Omega\big|_{r=r_{+}} = \frac{a}{\lambda\ell^{2}},
\end{aligned}
\end{equation}
where \(\Omega_{+}\) denotes the value of \(\Omega\) at the event horizon \(r = r_{+}\).

The diagonal form of the metric, Eq.~(\ref{DiagMetric}), is particularly useful for studying field dynamics, computing angular deficits, and evaluating thermodynamic quantities such as the entropy of the system. This representation considerably simplifies the analysis of the physical properties of the rotating black string.

\end{document}